\PassOptionsToPackage{}{graphicx}
\documentclass[fleqn]{2023SCGE}
\usepackage{graphicx}
\usepackage{multirow}
\usepackage{subfigure}
\usepackage{mleftright}

\newcommand*{\dif}{\mathop{}\!\mathrm{d}}
\newcommand*{\D}{\mathop{}\!\mathrm{D}}

\begin{document}
\ensubject{subject}
\ArticleType{Article}
\SpecialTopic{ }
\Year{}
\Month{}
\Vol{}
\No{}
\DOI{10.1007/}
\ArtNo{000000}
\ReceiveDate{}
\AcceptDate{}
\OnlineDate{}
\title{Probing Gravitational Quantum Field Theory through Polarization Fingerprints of Gravitational Waves}

\author[1,2,3]{Cong Xu}{xucong22@mails.ucas.ac.cn}
\author[3,4,5,6]{Hong-Bo Jin}{hbjin@bao.ac.cn}
\author[1,2,3,5,6]{Yue-Liang Wu}{ylwu@ucas.ac.cn}
\AuthorMark{Cong Xu}
\AuthorCitation{Cong Xu, H.-B. Jin and Yue-Liang Wu}
\address[1]{School of Fundamental Physics and Mathematical Sciences,Hangzhou Institute for Advanced Study, UCAS, Hangzhou 310024, China}
\address[2]{Institute of Theoretical Physics, Chinese Academy of Sciences, Beijing 100190, China}
\address[3]{University of Chinese Academy of Sciences, Beijing 100049, China}
\address[4]{National Astronomical Observatories, Chinese Academy of Sciences, Beijing 100101, China}
\address[5]{The International Centre for Theoretical Physics Asia-Pacific, University of Chinese Academy of Sciences, Beijing 100190, China}
\address[6]{Taiji Laboratory for Gravitational Wave Universe (Beijing/Hangzhou), UCAS, Beijing 100049, China}

\date{\today}
\abstract{
Gravitational Quantum Field Theory (GQFT) has been proposed as a candidate framework to reconcile general relativity with quantum field theory, and a distinctive imprint on gravitational-wave (GW) polarizations is crucially predicted. While general relativity allows only two tensor modes ($+, \times$), GQFT additionally favors a massless breathing scalar mode, providing a compelling yet largely unexplored observational target for testing quantum gravity. The central challenge is therefore to assess, in a mission-agnostic manner, how well future space-based interferometers can disentangle and detect these tensor and scalar polarization components across the sky. In this work, we develop a model-independent response formalism for LISA- and Taiji-like detectors by incorporating first-order orbital dynamics in the Solar System Barycenter frame. This framework yields three key observational consequences: (1) characteristic interference patterns between tensor and scalar modes, (2) a generalized, model-independent response function for the breathing mode, and (3) sky-position-dependent strategies that optimize detectability. We further translate the formalism into comprehensive polarization maps that provide complete sky coverage and remain fully compatible with existing mission designs, thereby circumventing the need for challenging direct breathing-mode measurements. Overall, our results deliver practical tools for future data analysis and establish a systematic avenue to test fundamental theories of gravity through their GW polarization fingerprints. }


\maketitle

\begin{multicols}{2} 
\section{Introduction}
Gravitational waves (GWs) have been detected by multiple experiments, and their polarizations play a crucial role in probing extended theories of gravity. In the framework of metric-compatible theories \Authorfootnote, there are up to six possible polarization modes \cite{Eardley1973}: two tensor modes ($+$ and $\times$), two vector modes ($x$ and $y$), and two scalar modes (breathing and longitudinal). For instance, in general relativity, only the plus and cross modes exist, which have been detected by the LIGO, Virgo, and KAGRA collaborations \cite{Abbott2023}. In Brans-Dicke gravity \cite{Brans1961,Brans1962}, an additional breathing mode arises due to the presence of a scalar field. Among various alternative theories, this paper focuses on Gravitational Quantum Field Theory (GQFT) \cite{Wu2015,Wu2016,Wu2023}.

GQFT was established to reconcile general relativity (GR) and quantum field theory (QFT). It is based on the fundamental principle that the laws of nature are determined by the intrinsic properties of matter's basic constituents. This principle distinguishes between two types of symmetries: intrinsic symmetries, which are defined by the quantum numbers of quantum fields as elementary particles, and external symmetries, which describe the motion of elementary particles in flat Minkowski spacetime\cite{Wu2015,Wu2016,Wu2023}. According to this framework, the intrinsic spin symmetry SP(1,3) must be localized into a spin gauge symmetry following the gauge symmetry principle.

In GQFT, the global Lorentz symmetry SO(1,3) in Minkowski spacetime and the intrinsic spin symmetry SP(1,3) in the Dirac fermion's spinor representation are unified into a joint symmetry structure SO(1,3)$\Join$SP(1,3). This framework replaces the conventional symmetry structure used in quantum field theory (QFT). To maintain these combined symmetries, we introduce a spin-related vector field $\hat{\chi}_{a}^{\; \mu}(x)$, which substitutes for the Kronecker symbol $\delta_{a}^{\;\mu}$ in QFT. This field exhibits bi-covariant transformation properties under both the spin gauge symmetry $SP(1,3)$ and the global Lorentz symmetry $SO(1,3)$, and it is treated as an invertible vector field. Its dual counterpart, $\chi_{\mu}^{\; a}(x)$, emerges as a spin-related gauge-type bi-covariant vector field. 

The gravigauge field $\chi_{\mu}^{\; a}$ behaves as a Goldstone-boson-type entity, corresponding to a massless graviton. Similar to how gravitational interactions occur for Dirac fermions, the coupling between the gravigauge field $\chi_{\mu}^{\;     a}(x)$ and the spin gauge field ${\mathcal A}_{\mu}^{ab}$ enables the construction of a spin-gauge-invariant action in GQFT. This structure cannot be replicated using the metric field alone. Importantly, the general linear group symmetry GL(4,$\mathbb{R}$), which is fundamental to GR, appears as an implicit symmetry within GQFT. A key distinction from GR is that GQFT maintains a flat Minkowski spacetime as its underlying spacetime framework. Interactions between the gravigauge field and the spin gauge fields and specified spinor fields introduce non-geometric effects, which result in violations of the equivalence principle observed in GR and give rise to novel physical phenomena \cite{Gao2024}. 

The gravigauge field $\chi_{\mu}^{\; a}$ of GQFT can be mathematically regarded as a frame field. 
Furthermore, the corresponding metric field can be expressed as $\chi_{\mu\nu}=\eta_{ab}\chi_{\mu}^{\; a}\chi_{\nu}^{\; b}$ traditionally used in GR. 
Meanwhile, GQFT can also be expressed geometrically as \cite{Wu2023}
\begin{gather}
    R_{\mu\nu} - \frac{1}{2} \chi_{\mu\nu} R + 8 \pi G_N \widetilde{T}_{\mu\nu} = 8\pi G_N T_{\mu\nu}, \nonumber \label{symEq} \\
    \bar{\nabla}_\rho \bar{F}^\rho_{[\mu\nu]} + \widetilde{T}_{[\mu\nu]} = m_G^{-2} T_{[\mu\nu]} \label{antiSymEq}. 
\end{gather}
Compared to GR, the first symmetric equation incorporates an additional term $\tilde{T}$, which the spin gauge field $A^a_\mu$ and the gravigauge field $\chi^a_\mu$ collectively provide. 
And the spin gauge field entirely dominates the second antisymmetric component. 
Although the theory fundamentally introduces the gravigauge field $\chi^\mu_a$ as a frame field and the spin gauge field $A^a_\mu$ to describe interactions, the field strength of the latter is rigorously decomposed into a homogeneous part and an inhomogeneous part. 
The homogeneous part is equivalently described by $\chi^\mu_a$, and its corresponding field strength is strictly the Riemann tensor associated with $\chi_{\mu\nu}$. 
Furthermore, under the classic situation, the spin gauge field within the symmetric energy-momentum tensor reduces exclusively to this homogeneous part, and the second antisymmetric equation acts as a constraint maintaining symmetry. 
Therefore, the gravitational equations can be treated in a metric form. 

Beyond its foundational implications, GQFT provides a theoretical framework for exploring a hyperunified field theory capable of unifying all fundamental interactions \cite{Wu2017, Wu2022}. Recently, a general theory of the Standard Model has been developed within the GQFT framework. This theory integrates the Standard Model of particle physics and the standard cosmological model, offering new insights into the mysteries of the universe's dark sector. Specifically, it provides fresh perspectives on understanding the nature of dark matter, the dynamics of the early inflationary universe, and the role of dark energy in the current cosmic expansion.

Numerous studies have been conducted within the framework of GQFT, addressing topics such as inflation \cite{Wang2023}, dark matter \cite{Wang2022}, and particle physics \cite{Huang2023}. The Taiji mission, a space-based gravitational wave (GW) observatory, has been analyzed within the parametrized post-Einsteinian framework to evaluate its capability to detect different polarization modes. Studies indicate that Taiji can measure both dipole and quadrupole GW emissions \cite{TaijiScientific:2021qgx, Gong:2021gvw, Liu:2020mab}. Similarly, the LISA-Tianqin network has been investigated for its sensitivity to the polarization modes of the stochastic GW background \cite{Hu:2024toa}. In a general four-dimensional metric theory of gravity, the detection of additional polarization modes beyond the two predicted by GR would suggest the need for an extended theory of gravitation. The specific polarizations observed could help rule out certain theoretical models \cite{Nishizawa:2009bf}.
Both Taiji \cite{Hu2017} and LISA \cite{Danzmann1996} employ a triangular three-detector configuration with equal arm lengths. This paper focuses on the polarization modes in GQFT and their corresponding response characteristics in such a three-detector setup.
 
The interaction of gravitational waves (GWs) with a detector arm is manifested as fluctuations in the arm length between two detectors in a triangular three-detector configuration. The intensity of the detector's response depends on the angle between the GW propagation direction and the detector arm. This response is minimal when the GW propagation direction is parallel to the arm and reaches its maximum when the GW direction is perpendicular to the arm. 
Since gravitational wave sources are distributed across the sky, the angle between the GW propagation direction and the detector arm is not constant. This variation arises due to the orbital motion of space-based detectors, such as Taiji \cite{Hu2017} and LISA \cite{Danzmann1996}, around the Sun. As these detectors continuously change their positions and orientations, the angle between the GW propagation direction and the detector arm fluctuates, leading to variations in the detector's sensitivity to GW signals.

Laser interferometers detect gravitational waves (GWs) by measuring the differential length changes between two detector arms. However, maintaining perfectly equal arm lengths over the entire orbital period of a space-based detector is impractical. This challenge necessitates the use of Time Delay Interferometry (TDI) to account for varying arm lengths. 
Recent advancements in atomic clock technology offer an alternative approach. By precisely measuring the optical path time difference between detectors, these clocks can infer arm-length variations without relying solely on TDI. This method has sparked significant interest in the development of high-precision optical clocks for space-based GW detection\cite{Kolkowitz:2016wyg,Tino:2019tkb,He:2020elt,Wang:2024tnk}.

In this paper, we concentrate on investigating novel polarization modes predicted by GQFT that extend beyond GR. Our analysis specifically examines the response of individual detector arms to gravitational wave signals, and does not consider dual-arm interferometric measurements to maintain analytical focus. The paper is organized as follows: we first introduce the polarization modes in metric theories of gravity in \S{\ref{polSec}}, then derive the specific polarization modes within the GQFT framework in \S{\ref{GQFT}}. We further present the detector response characteristics in the Solar System Barycenter frame in \S{\ref{response}}, provide a universal, model-independent response function for the Taiji configuration, and identify the optimal observational timing windows for GQFT in \S{\ref{distinction}}.

\section{Methods}            
\subsection{Polarization in metric theory}\label{polSec}

The action of GW on test masses can be understood through their tidal effects, which manifest as deviations from geodesic motion. This phenomenon is captured by the Jacobi field equation for geodesic deviation,
$\dfrac{\D^2 X^\alpha}{\dif \tau^2}=T^\gamma \nabla_\gamma (T^\delta \nabla_\delta X^\alpha)=-R^\alpha_{\beta \delta \gamma} X^\delta T^\gamma T^\beta.$ In metric theories of gravity, the polarization state is entirely determined by the electric components of the Riemann tensor \cite{Dadhich2000}. 
When examined in the gravitationally independent flat frame (the natural framework for GQFT), this description simplifies to the form, $\dfrac{\dif^2 X^i}{\dif \tau^2}=-R^i_{0j0} X^j$, which describes weak, plane, null gravitational waves \cite{Eardley1973}. The symmetry properties of the Riemann tensor reveal that $R^i_{0j0}$ possesses six independent components, indicating that gravitational waves can exhibit up to six distinct polarization modes.

For a wave propagating along the z-direction, we define $R_{i0j0}(t)$ as the ``driving force matrix'' $S_{ij}(t)$, a 3-dimensional symmetric matrix. The polarization modes can be further classified using the Newman-Penrose formalism \cite{Newman1962}. Working with tetrad indices $(l,n,m,\bar{m})$, we consider tensor contractions with the corresponding null vectors, $X_{efgh} := X_{\mu \nu \rho \sigma} e^\mu f^\nu g^\rho h^\sigma$, where $(e,f,g,h)$ range over the null tetrad basis $(l,n,m,\bar{m})$. Taking a null tetrad frame as follows:
\begin{align}
        l&=\frac{1}{\sqrt{2}} (\partial_t + \partial_z), \quad n=\frac{1}{\sqrt{2}} (\partial_t - \partial_z)\nonumber\\
        m&=\frac{1}{\sqrt{2}} (\partial_x + i \partial_y), \quad \bar{m}=\frac{1}{\sqrt{2}} (\partial_x - i \partial_y).
\end{align}
In the null frame, where the field depends solely on retarded time $(t-z)$, the Riemann tensor satisfies the condition $R_{efgh,p}=0$. Here, $(e,f,g,h)$ span the null tetrad $(l,n,m,\bar{m})$, while $(p,q)$ range over $(l,m,\bar{m})$. Substituting these into the differential Bianchi identity yields $R_{ef[pq,n]}=\dfrac{1}{3} R_{efpq,n}=0$. This implies that, up to trivial constants, $R_{efpq}=R_{pqef}=0$. Consequently, the only non-vanishing components of the Riemann tensor are of the form $R_{pnqn}$.

Under the null condition, the original 12 NP scalars reduce to just 4 independent quantities: $\Psi_2,\Psi_3,\Psi_4,\Phi_{22}$: two real scalars $\Psi_2,\Phi_{22}$ and two complex scalars $\Psi_3,\Psi_4$. These remaining NP null scalars collectively contain 6 independent degrees of freedom, corresponding exactly to the degrees of freedom in the Riemann tensor for weak, plane, null gravitational waves \cite{Hyun2019}:
\begin{align}
        \Psi_2 &= \dfrac{1}{6} R_{3030},\quad \Phi_{22} = R_{1010} + R_{2020},\quad \Re(\Psi_3) = \dfrac{1}{2} R_{3010},\nonumber\\
         \Im(\Psi_3) &= -\dfrac{1}{2} R_{3020},\quad 
        \Re(\Psi_4) = R_{1010} - R_{2020},\quad  \Im(\Psi_4) = -2 R_{1020}.
\end{align}
where $\Re$ and $\Im$ represent the real and imaginary parts, respectively. 
Then, the driving force matrix can be expanded into a linear combination of polarization bases.
\begin{equation}
    S(t) = \sum_A p_A(e_z,t) E_A(e_z),
\end{equation}
where $A$ ranges over $(+,\times,x,y,b,l)$, covering six polarization modes. 

In the above expansion, the following bases of polarization matrices are adopted:
\begin{align}
 E_+(e_z)&:=
        \begin{bmatrix}
            1 & 0 & 0 \\
            0 & -1 & 0 \\
            0 & 0 & 0 
        \end{bmatrix},
        \quad
        E_\times (e_z):=
        \begin{bmatrix}
            0 & 1 & 0 \\
            1 & 0 & 0 \\
            0 & 0 & 0 
        \end{bmatrix},
        \quad
        E_x(e_z):=
        \begin{bmatrix}
            0 & 0 & 1 \\
            0 & 0 & 0 \\
            1 & 0 & 0 
        \end{bmatrix},\nonumber \\
        E_y(e_z)&:=
        \begin{bmatrix}
            0 & 0 & 0 \\
            0 & 0 & 1 \\
            0 & 1 & 0 
        \end{bmatrix},
                \quad
               E_b(e_z):=
        \begin{bmatrix}
            1 & 0 & 0 \\
            0 & 1 & 0 \\
            0 & 0 & 0 
        \end{bmatrix},
        \quad
        E_l(e_z):= \sqrt{2}
        \begin{bmatrix}
            0 & 0 & 0 \\
            0 & 0 & 0 \\
            0 & 0 & 1 
        \end{bmatrix}, 
\end{align}
where the basis is chosen to be normalized to 2. This normalization condition leads to corresponding polarization amplitudes:
\begin{align}
        p_l(e_z,t)&=3\sqrt{2} \Psi_2 =\frac{1}{\sqrt{2}} R_{3030},\nonumber \\
        p_x(e_z,t)&=2 \Re(\Psi_3) = R_{3010},\nonumber \\
        p_y(e_z,t)&=-2 \Im(\Psi_3) = R_{3020},\nonumber \\ 
        p_+(e_z,t)&=\frac{1}{2} \Re(\Psi_4) = \frac{1}{2}(R_{1010} - R_{2020}),\nonumber \\
        p_\times (e_z,t)&=-\frac{1}{2} \Im(\Psi_4) = R_{1020},\nonumber \\
        p_b(e_z,t)&=\frac{1}{2} \Phi_{22} = \frac{1}{2}(R_{1010} + R_{2020}). \label{allOfPol}
\end{align}

\subsection{ Polarization in GQFT}\label{GQFT}

As stated in the introduction, when the spin gauge field decouples in the classical regime, its corresponding contribution can be fully described by the metric field associated with the gravigauge field, which justifies the extraction of polarization modes through standard geodesic deviation equations. 
In this framework, the metric is constructed from the gravigauge field $\chi^a_\mu$ as $\chi_{\mu\nu} =  \chi^a_\mu \chi^b_\nu \eta_{ab}$. In general, the gravigauge field can always be written as $\chi_{\mu}^{\; a}  \equiv \eta_{\mu}^{\; a} + h_{\mu}^{\; a}/2$. The corresponding metric takes the form $\chi_{\mu\nu} = \eta_{\mu\nu} + h_{(\mu\nu)}$, with $h_{(\mu\nu)} \equiv (h_{\mu\nu} + h_{\nu\mu})/2 + h_{\mu}^{\; a}h_{\nu}^{\; a}\eta_{ab}/4$. Here, $h_{(\mu\nu)}$ is treated using scalar-vector-tensor decomposition. 

Dynamical analysis of GQFT identifies five fundamental degrees of freedom in the gravitational sector \cite{Gao2024}. Although this implies the theoretical existence of five gravitational wave polarization modes, only three physical polarizations emerge as observationally significant when analyzing their tidal effects on test masses. The geodesic deviation equation, which governs observable gravitational wave effects, exclusively captures tidal forces. Consequently, the remaining two vector-mode polarizations remain observationally inaccessible within this framework, as they encode intrinsic spin properties and represent manifestations of equivalence principle violations rather than measurable tidal deformations.

For a plane GW propagating along the z-direction, it acts as a perturbation of the metric, which implies that derivatives in the $x$--$y$ directions vanish. The spin-2 sector reproduces the standard transverse-traceless modes of GR as follows:
\begin{equation}
    \begin{bmatrix}
        ~ & ~              & ~              & ~\\
        ~ & \hat{h}_+      & \hat{h}_\times & ~\\
        ~ & \hat{h}_\times & -\hat{h}_+     & ~\\
        ~ & ~              & ~              & ~
    \end{bmatrix} \label{spin-2} ,
\end{equation}
whilst satisfying $\Box \hat{h}_{ij}=0$.

The spin-1 components of the metric perturbation follow the decomposition: $h_{it} = S_i, h_{ij} = 2 \partial_{(i} F_{i)}$. GQFT dynamics establish the relation between $S_i$ and $F_i$ as $S_i=\partial_t F_i$.
Meanwhile, there is the transverse condition  $\partial_i F^i =0$ for the vector field, implying the traceless condition of the corresponding tensor $h_{ij}$. For plane waves propagating along the z-direction, these constraints reduce the degrees of freedom to just $F_1$ and $F_2$. The resulting spin-1 sector can therefore be expressed as follows:
\begin{equation}
    \begin{bmatrix}
        ~      & \partial_t F_1 & \partial_t F_2 & ~\\
        \partial_t F_1 & ~              & ~              & \partial_z F_1\\
        \partial_t F_2 & ~              & ~              & \partial_z F_2\\
        ~      & \partial_z F_1         & \partial_z F_2         & ~
    \end{bmatrix} \label{spin-1} ,
\end{equation}
whilst satisfying $\Box F_i=0.$

The spin-0 components of the metric perturbation are given by: $h_{tt}=0, h_{it}=-\partial_i B, h_{ij} = \delta_{ij} \mleft(-2 \psi \mright) + 2 \partial_i \partial_j E$. GQFT introduces two gauge-invariant variables: $\Phi:=\varphi - \dfrac{1}{2} \partial_t B$ and $A := B + 2 \partial_t E$. The resulting spin-0 part along the z-direction is given by
\begin{equation}
    \begin{bmatrix}
    -2 \varphi & ~ & ~ & -\partial_z B\\
    ~ & -2 \psi & ~              & ~\\
    ~ & ~              & -2 \psi & ~\\
    -\partial_z B & ~         & ~         & -2 \psi + 2 \partial_z^2 E
    \end{bmatrix} \label{spin-0} .
\end{equation}
GQFT dynamics establish the relations: $\psi = - \dfrac{1}{2} \gamma_W \Phi$ and $\partial_t A = -(\gamma_W - 2) \Phi$. All scalar components satisfy the wave equation. The full metric perturbation $h$ is obtained by adding up spin-0 \eqref{spin-0}, spin-1 \eqref{spin-1}, and spin-2 \eqref{spin-2} components.

This enables the calculation of GW polarization amplitudes in GQFT through the described procedure based on Equation~\eqref{allOfPol}.
\begin{align}
        p_+ &= -\frac{1}{2} \partial_t^2 h_+, \quad  
        p_\times = -\frac{1}{2} \partial_t^2 h_\times,\nonumber \\
        p_x &= - \frac{1}{2} (\partial_t^2 h_{13} - \partial_t \partial_z h_{01}) \equiv 0,  \quad 
        p_y = - \frac{1}{2} (\partial_t^2 h_{23} - \partial_t \partial_z h_{02}) \equiv 0, \nonumber \\
        p_b &= \partial_t^2 \psi,  \quad 
        p_l = \frac{1}{\sqrt{2}} (\partial_z^2 \varphi - \partial_t \partial_z^2 B + \partial_t^2 \psi - \partial_t^2 \partial_z^2 E) \equiv 0.
\end{align}
Here, the plus (+) and cross ($\times$) polarization modes in GQFT exactly match those in GR. However, the x and y vector modes completely vanish due to the constraints for the spatial components: $h_{13} = \partial_z F_1, h_{23} = \partial_z F_2$, and the temporal components: $h_{0i} = S_i = \partial_t F_i$ in GQFT, which together eliminate the spin-1 degrees of freedom from contributing to observable polarizations. Similarly, the longitudinal mode vanishes entirely when considering the gauge-invariant variables, $\Phi:=\varphi - \dfrac{1}{2} \partial_t B, A := B + 2 \partial_t E$, the scalar field relation, $\psi = - \dfrac{1}{2} \gamma_W \Phi$, the constraint equation, $\partial_t A = -(\gamma_W - 2) \Phi$, as well as the wave equation along the z-direction, $\Box \psi =0$. Consequently, three of the five polarization modes predicted in GQFT become observable through tidal deformations: two transverse-traceless tensor modes ($+$, $\times$) as in GR, and one additional breathing mode induced by $\psi$.

It is natural here to compare with generic scalar-tensor theories, with the addition of a scalar field, scalar-tensor theories contain an extra breathing and longitudinal mixed mode. 
When the parameter $\omega$ is a constant, the theory reduces to Brans-Dicke theory, which retains only an extra breathing mode. Geometrically, Brans-Dicke theory and GQFT even share the same electric components of the Riemann tensor, though their origins and strengths may differ. 
In Brans-Dicke-type scalar-tensor theories, the framework explicitly introduces an additional scalar field, and it is natural for scalar modes to be induced. However, GQFT is not supplemented with any auxiliary fields. 
Instead, the intrinsic Goldstone-boson-like properties of the gravigauge field inherently suppress the longitudinal modes, which consequently preserves only the transverse scalar breathing mode. 
On the other hand, the breathing mode in GQFT does not purely originate from conventional geometric contributions. 
Rather it is significantly driven by an additional antisymmetric source that the spin gauge field provides \cite{gao2025}, which inherently causes the physical intensities to differ. 

A comprehensive testing framework for GQFT requires extending beyond traditional geodesic deviation measurements to include spin-gravity coupling effects between test particles and GWs, possible spin-dependent modulation of detector responses, and novel detection signatures in spin-polarized measurement systems. This expanded approach may uncover observable manifestations of the currently hidden spin-1 degrees of freedom, potentially revealing spin-polarization correlations in wave detection, frequency-dependent modulation effects, and novel torsion-like couplings in the detector response.

\subsection{Response on the detectors}\label{response}

As mentioned above, GW propagation can be expressed as a linear combination of different polarization tensors \cite{Nishizawa2009}:
\begin{equation}
    \mathbf{h} (t) = h_+(t) \varepsilon_+ + h_\times (t) \varepsilon_\times + h_x(t) \varepsilon_x + h_y(t) \varepsilon_y + h_b(t) \varepsilon_b + h_l(t) \varepsilon_l \label{pol}
\end{equation}
We first consider a GW propagating along $\hat{k}$. By adopting a coordinate system defined by the basis $\{\hat{k},\hat{m},\hat{n}\}$, the corresponding polarization basis can be constructed as follows:
\begin{align}
        \varepsilon_+ &= \hat{m} \otimes \hat{m} -\hat{n} \otimes \hat{n}, \quad \varepsilon_\times = \hat{m} \otimes \hat{n} +\hat{n} \otimes \hat{m},\nonumber \\
        \varepsilon_x &= \hat{m} \otimes \hat{k} +\hat{k} \otimes \hat{m}, \quad \varepsilon_y = \hat{n} \otimes \hat{k} +\hat{k} \otimes \hat{n},\nonumber \\
        \varepsilon_b &= \hat{m} \otimes \hat{m} +\hat{n} \otimes \hat{n}, \quad \varepsilon_l = \sqrt{2} \hat{k} \otimes \hat{k}. \label{polBasis}
\end{align}

However, actual detection occurs in the Solar System Barycenter (SSB) frame, where the corresponding coordinate system is denoted by $\{\hat{x},\hat{y},\hat{z}\}$. In this frame, the polarizations require additional projection and mixing compared to the source coordinate system.

Mathematically, these two coordinate systems are related by a rotation parameterized by Euler angles that form a chart on $SO(3)$. Their physical interpretations differ: $\{\theta,\varphi\}$ correspond to the sky location and specify the GW propagation direction, while $\psi$ is the polarization angle that describes the polarization basis and its mixing. Thus, in the SSB frame $\{\hat{x},\hat{y},\hat{z}\}$, the GW direction is typically described using the basis $\{\hat{k},\hat{u},\hat{v}\}$:
\begin{align}
        \hat{k} &= (\sin \theta \cos \varphi,\sin \theta \sin \varphi ,\cos \theta ),\nonumber\\
        \hat{u} &= (\cos \theta \cos \varphi,\cos \theta \sin \varphi ,-\sin \theta ),\nonumber\\
        \hat{v} &= (-\sin \varphi,\cos \varphi,0).
\end{align}
This corresponds to a rotation by $\psi$ relative to the source coordinate system $\{\hat{k},\hat{m},\hat{n}\}$. The polarization basis vectors are then projected as:
\begin{align}
        \hat{m} &= \cos \psi \hat{u} + \sin \psi \hat{v},\nonumber\\
        \hat{n} &= -\sin \psi \hat{u} + \cos \psi \hat{v}.
\end{align}
This projection enables us to consistently treat both the GW polarization response and detector orbital motion within the SSB reference frame. 

Nearly monochromatic gravitational waves are generated under the quadrupole approximation \cite{peters1963gravitational}.
For HM Cancri, we refer to the waveform presented in \cite{Jin:2024mma,korol2022observationally}.
\begin{align}
        h_+(t) &= \frac{2 (G \mathcal{M})^{5/3} (\pi f)^{2/3}}{d_L c^4} (1 + \cos^2 \iota) \cos \Phi(t), \\
        h_\times (t) &= \frac{4 (G \mathcal{M})^{5/3} (\pi f)^{2/3}}{d_L c^4} \cos \iota \sin \Phi(t).
\end{align}
The treatment of the breathing mode follows the same procedure as for the tensor modes. 
A comparative analysis based on tensor mode GQFT wave equations reveals that the $b$-mode and tensor modes differ by a characteristic factor $\frac{(1+\gamma_W)}{(1-\gamma_W/2)}$, 
where the parameter $\gamma_W$ from the coupling constants $\alpha_G$, $\alpha_W$ and $\gamma_G = 8 \pi G m_G^2$ within GQFT $\gamma_W = \gamma_G (\alpha_G - \alpha_W/2)$, which $\alpha_G$ and $\alpha_W$ utilize to construct the tensor $\bar{\eta}$ in Equation~(82) \cite{Wu2023}, and is subject to physical constraints on $(2.1 \pm 2.3) \times 10^{-5}$ \cite{Gao2024}. 

Indeed, the contributions to the breathing mode in GQFT are not solely from the geometric source, but also receive significant contributions from the spin gauge field \cite{gao2025}
\begin{equation}
    \Psi(t,x) = - \frac{G_N M}{(1-\gamma_W)r} + \frac{G_N}{(1-\gamma_W)r} \mleft[ \frac{\gamma_W}{6} \frac{\dif^2}{\dif t_r^2} \mathcal{I}^k_k - \frac{2}{3} (1 + \gamma_W) \frac{\dif}{\dif t_r} \tilde{S} \mright], 
\end{equation}
here clearly indicate that two distinct contributors to the breathing mode exist. 
One contributor is the normal quadrupole, whereas the other is an antisymmetric source $\tilde{S}$, which the spin gauge field intrinsically provides.
Specifically, the antisymmetric source scales as $\sim 1/m_G^2$, while the quadrupole component scales as $\sim G_N \sim 1/m_P^2$. 
Where $m_G$ is the fundamental scale characterizing the typical masses of spin gauge fields obtained by breaking the scale symmetry in the GQFT, and $m_P$ is the Planck scale. 
So it evident that the contribution from the spin gauge sector in GQFT substantially exceeds the traditional geometric quadrupole contribution. 
Meanwhile $1/m_G^2$ cancels out with the $m_G^2$ coefficient present in the Equation~\eqref{symEq} of GQFT with corresponding weak field
\begin{equation}
    R_{\mu\nu} - \frac{1}{2} \chi_{\mu\nu} R + 8 \pi G_N m_G^2 \widetilde{R}_{\mu\nu} = 8\pi G_N T_{\mu\nu}, 
\end{equation}
where $\widetilde{R}$ is the $\widetilde{T}$ retaining only the homogeneous contribution of the spin gauge field in the weak-field regime. 
Therefore, unlike conventional pure geometric breathing mode being very weak, the GQFT breathing mode amplitude can be comparable to tensor modes. 

But even so, the breathing mode may still be difficult to observe with ground-based detectors. 
For example, the LIGO detector operates as an L-shaped Michelson interferometer, which primarily optimize for the observation of tensor modes. 
Due to the breathing mode is entirely isotropic, that the observable strain of the breathing mode becomes exactly zero when the tensor mode reaches its maximum measurable intensity. 
This physical phenomenon occurs that the detector exhibits maximum response when the propagation direction of the gravitational wave is orthogonal to the arms, consequently, the interference operation cancels the breathing mode. 
For the breathing mode observations, it is necessary to select observations with different angles and relative positions, space-based detectors orbiting the sun are far more likely to observe this unique polarization mode. 

Under the specific physical conditions of a nearly monochromatic chirp signal, the waveform is established to be nearly model-independent, therefore the primary theoretical discrepancies between theories manifest directly within the polarization modes themselves. 
Consequently, the paper adopt a simplified waveform accommodating the breathing mode with common tensor modes, which provides a universal reference that facilitates the distinction of polarization signatures without relying on highly model-dependent waveform generations. 
Setting the orbital inclination and initial phase to zero also helps maintain consistency in the waveform amplitudes. We employ the following waveform:
\begin{align}
        h_+(t) &= \frac{4 (G \mathcal{M})^{5/3} (\pi f)^{2/3}}{d_L c^4} \cos \Phi(t), \\
        h_\times (t) &= \frac{4 (G \mathcal{M})^{5/3} (\pi f)^{2/3}}{d_L c^4} \sin \Phi(t), \\
        h_b(t) &= \frac{2 (1+\gamma_W) (G \mathcal{M})^{5/3} (\pi f)^{2/3}}{(1-\gamma_W/2) d_L c^4} \cos \Phi(t).
\end{align}
Here, $d_L$ is the luminosity distance, $\mathcal{M} = M \eta^{\frac{3}{5}}$ is the chirp mass with $\eta = \dfrac{m_1 m_2}{M^2}$ and $M = m_1 + m_2$ is the total mass. 
Meanwhile, for the phase, we consider a near-monochromatic chirp signal, i.e., $\Phi(t)=2\pi f t+ \pi \dot{f} t^2 +\phi_0$, where $\dot{f} =\dfrac{96}{5} \pi^{8/3} \left( \dfrac{G \mathcal{M}}{c^3}\right)^{5/3} f^{11/3}. $
Since this work aims to develop a polarization-focused model for distinguishing between GQFT and GR, the analysis remains independent of specific waveform models, thereby maintaining generality.

Turning to detector trajectories, both the LISA and Taiji missions follow Keplerian orbits and maintain an arm length of $2.5$--$3$ million kilometers. To first order in eccentricity, the spacecraft positions in SSB coordinates can be expressed as \cite{Rubbo2004}:
\begin{align}
        x(t) &= R \cos \alpha + \frac{1}{2} e R \bigl(\cos (2 \alpha - \beta) - 3 \cos \beta \bigr) \\ 
        y(t) &= R \sin \alpha + \frac{1}{2} e R \bigl(\sin (2 \alpha - \beta) - 3 \sin \beta \bigr) \\ 
        z(t) &= - \sqrt{3} e R \cos (\alpha - \beta) \label{orbit} ,
\end{align}
where $R = 1 \mathrm{AU}$ represents the radial distance to the guiding center, $e = \dfrac{L}{2 \sqrt{3} R}$ denotes the eccentricity, $\alpha = 2 \pi f_m t + \kappa $ is the orbital phase of the guiding center, and $\beta = \dfrac{2 \pi n}{3} + \lambda (n = 0,1,2)$ describes the relative phase of the spacecraft within the constellation. Setting $\kappa = \lambda = 0$ corresponds to the initial ecliptic longitude and orientation of the constellation.

When examining the effect of GWs on the detector, we consider a photon emitted from spacecraft $i$ at time $t_i$ and received at spacecraft $j$ at time $t_j$. The nominal arm length is $\ell_{ij}$, and its variation under low-frequency GWs (where the transfer function approaches unity) can be expressed as
\cite{Rubbo2004,Jin:2024mma,Guo2024}
\begin{equation}
    \delta \ell_{ij} = \frac{1}{2} \hat{r}_{ij}^a (t) \otimes \hat{r}_{ij}^b (t) \int^{t_j}_{t_i} \mathbf{h}_{ab}(t - \hat{k}\cdot \mathbf{x}(t)) \dif t . \label{strain}
\end{equation}
Here, $a,b$ are abstract indices denoting tensor contractions, and $\hat{r}_{ij}^a(t) = \dfrac{\mathbf{x}^a_j(t_j) - \mathbf{x}^a_i(t_i)}{\ell_{ij}(t_i)}$ represents the unit vector along the arm direction derived from the orbit (see Equation~\eqref{orbit}).

While interferometric observables typically measure differential changes between adjacent arms, our analysis focuses on the distinctive breathing mode in GQFT, which is transverse and isotropic. Since differential combinations could potentially suppress this mode's signature, we instead examine single-arm responses through the length strain itself.
This approach, combined with a metric perturbation expansion by polarization components, enables isolated examination of each polarization’s contribution.

Furthermore, Equation~\eqref{strain} can be recast by substituting Equation~\eqref{pol}, which shows that the length strain receives contributions from both the waveform amplitude $A:= \int h(t)\,\dif t$ and the response function $F := \frac{1}{2}\,\hat{r}_{ij}^a(t)\otimes \hat{r}_{ij}^b(t)\,\varepsilon_{ab}$. Thus, the length strain is a linear combination of the amplitude and the response function, $\delta\ell/\ell = \sum_{i\in \mathrm{pol}} A_i F_i$. All model dependence is contained in the waveform, while the response function is decoupled from the specific waveform and depends solely on the detector configuration and position.

Therefore, by providing a model-independent response function, we can effectively characterize the response of different polarization modes in the Taiji orbit.
Thereafter, any waveform model simply provides the amplitude, which is then linearly combined with the corresponding polarization response function to yield the length strain. To clearly illustrate the performance of each polarization mode, we neglect polarization mixing by setting $\psi=0$; see Supplementary Materials for the analytical expression for the response function in an arbitrary direction.

\begin{figure*}
\includegraphics[width=\textwidth]{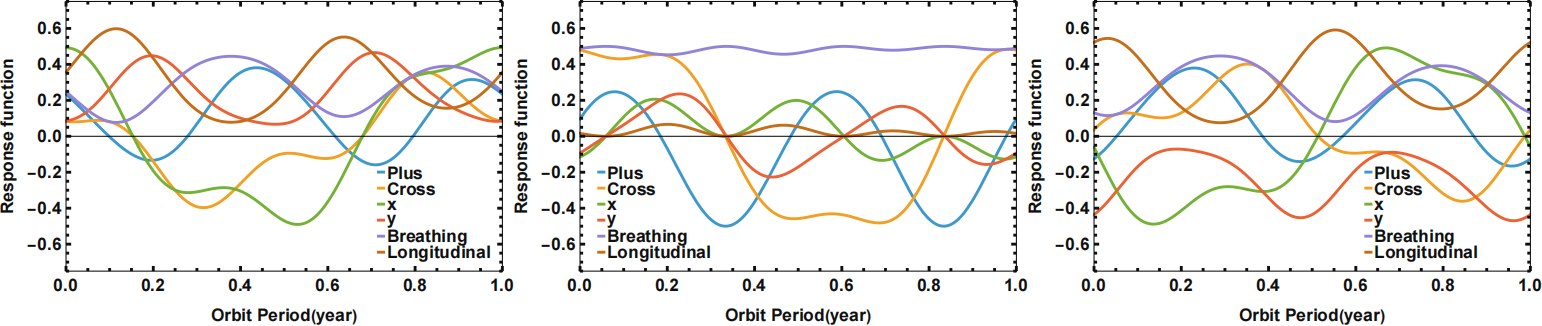}
\caption{Response functions at sky position $(\theta, \varphi)=(1.65, 2.10)$ radians for arms 12, 13, and 23. Blue, orange, green, red, purple, and brown correspond to the $+$, $\times$, $x$, $y$, breathing, and longitudinal responses, respectively.}
\label{responseFunc}
\end{figure*}

Specifically, we provide plots of the response $F$ for different detector arms. 
Here, we take HM Cancri (RX J0806.3+1527) as an example GW source, located at sky position $(\theta, \varphi)=(1.65, 2.10)$ radians.
At this point, the response functions for all polarization modes on different arms can be plotted as shown in Figure~\ref{responseFunc}.

The response function demonstrates the impact of the Taiji orbit on gravitational wave signals. By comparing the three detector arms, we find that the tensor modes exhibit a relatively stable periodic variation, taking both positive and negative values across all arms. In contrast, the vector modes show significant arm-to-arm variations in their projections. Specifically, the $y$ mode can be entirely positive (or entirely negative) over certain time intervals, indicating that the vector projection is highly sensitive to the detector's orientation. Finally, the scalar modes are consistently positive and regular regardless of the detector arm, reflecting the fact that scalar responses are largely insensitive to the arm projection.

\section{Results in GQFT}\label{distinction}

We consider a GW source as a nearly monochromatic chirp signal, as generated by compact binaries such as double white dwarfs. As an illustrative example, we adopt HM Cancri (RX J0806.3+1527) with sky location $(\theta, \varphi)=(1.65, 2.10)$ radians, component masses $0.55\,\mathrm{M}_\odot$ and $0.27\,\mathrm{M}_\odot$, GW frequency $6.22\,\mathrm{mHz}$, and luminosity distance $5\,\mathrm{kpc}$. 

The detector response depends on the relative orientation between the detector arm and the GW source; length changes provide a convenient description of the signal.
This allows us to separately analyze the responses to plus, cross, and breathing polarization modes over a one-year observation period, and to quantify the differences between GR and GQFT predictions.

Given the vast disparity in timescales—the source's characteristic period being on the order of seconds, while the response function's period is on the order of years—the resulting signal in the time domain is highly oscillatory. We employ conventional time-frequency analysis methods for signal extraction. As the chirp signal in this context is nearly monochromatic, we can effectively analyze the distinctions between different gravitational theories by plotting a slice of the corresponding spectrogram. 

\begin{figure*}
    \includegraphics[width=\textwidth]{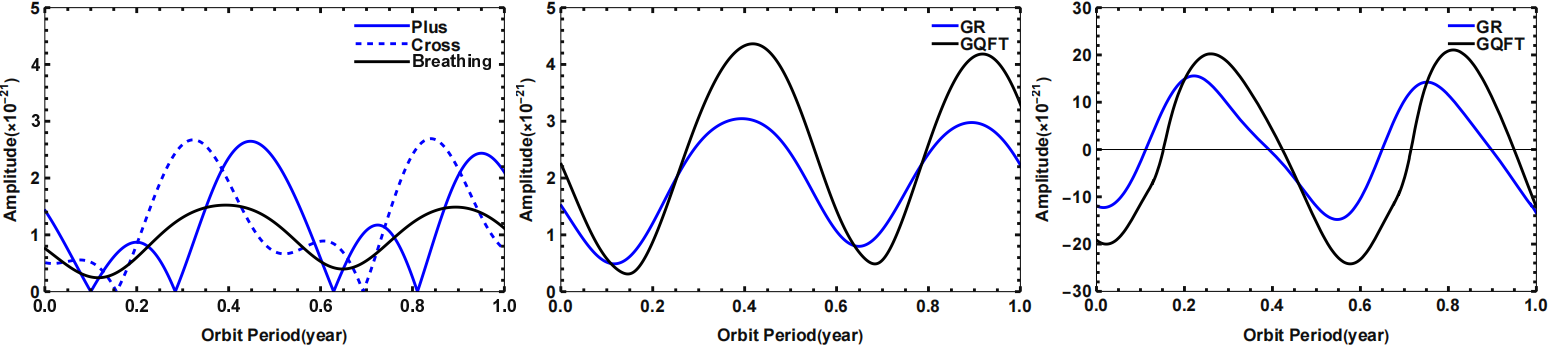} 
    \caption{GW strain response of arm12 to HM Cancri at a specific sky location. Left panel: length-strain components for the $+$, $\times$, and breathing modes. Middle panel: total strain comparison between GR (mixture of $+$ and $\times$) and GQFT (including all three polarizations). Right panel: relative temporal rate of change in the strain amplitude for GR and GQFT, shown as a supplementary reference for the middle panel.}
    \label{dif12}
\end{figure*}

\begin{figure*}
    \includegraphics[width=\textwidth]{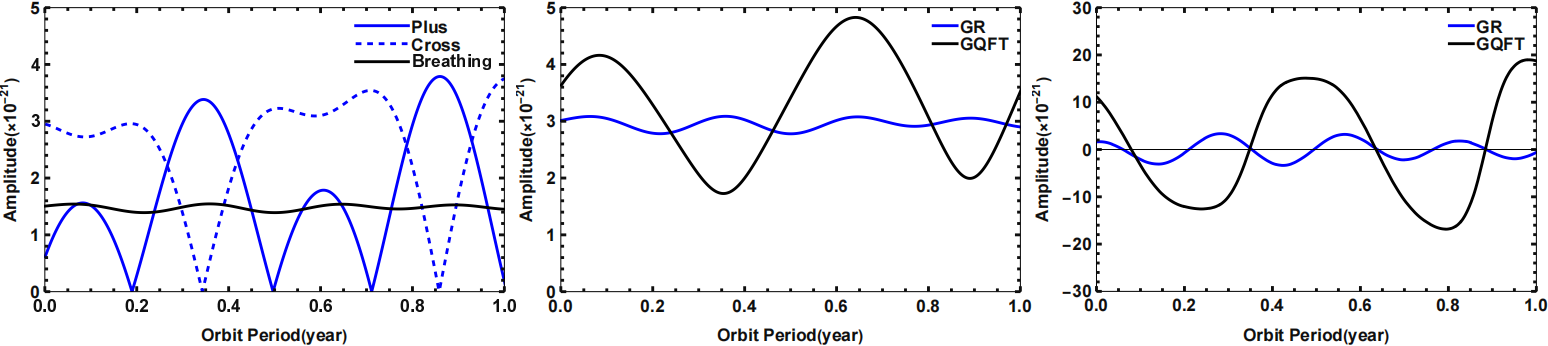} 
    \caption{Same as Figure~\ref{dif12}, but for arm13.}
    \label{dif13}
\end{figure*}

\begin{figure*}
     \includegraphics[width=\textwidth]{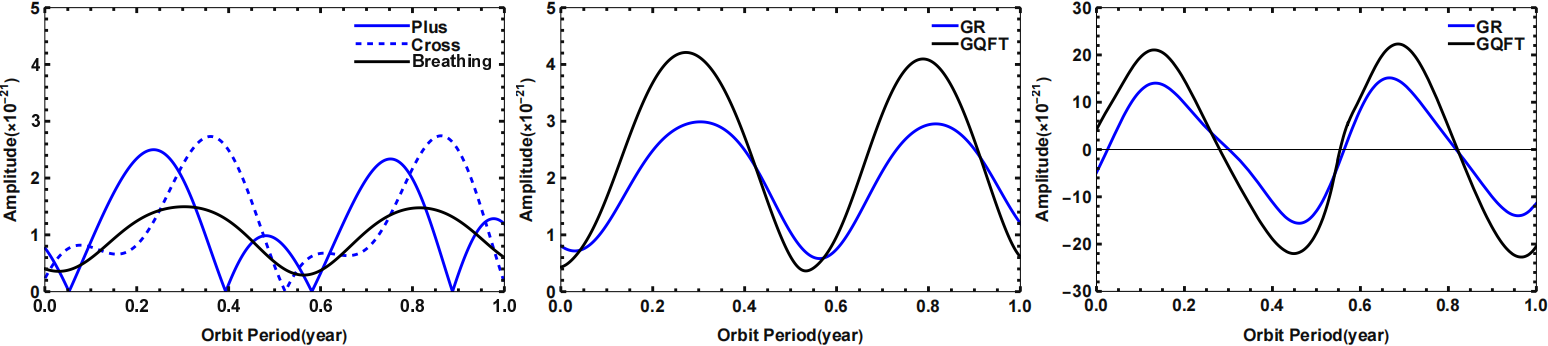} 
    \caption{Same as Figure~\ref{dif12}, but for arm23.}
    \label{dif23}
\end{figure*}

The distinctions observed in Figures~\ref{dif12}--\ref{dif23} can be traced to the relative source--detector orientation. In particular, arm13 exhibits a notable behavior compared to the other two arms: along this arm, the plus and cross polarizations show a marked symmetry in their distributions, while the $b$-mode fluctuations remain small. Consequently, the total strain predicted by GR and GQFT can differ more strongly in amplitude. This suggests that, for RX~J0806.3+1527, arm13 offers enhanced capability to probe the $b$ mode and to distinguish between GR and GQFT. 

\begin{figure*}
    \includegraphics[width=\textwidth]{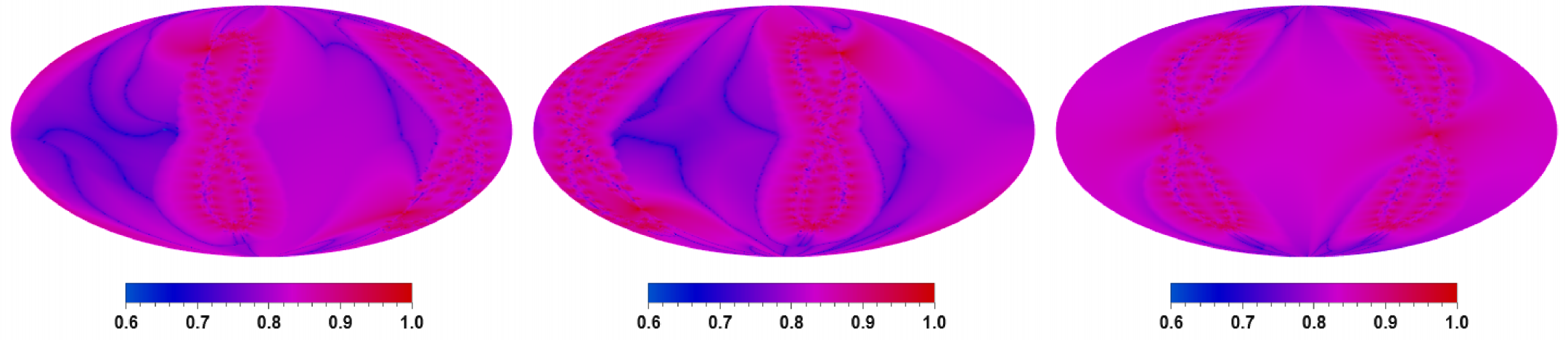} 
    \caption{All-sky maps illustrating the logarithmic difference in the maximum-to-minimum amplitude ratios between GQFT and GR for arm12, arm13 and arm23 (from left to right), to effectively visualize the non-linear dynamic range of these discrepancies across the sky. 
    Redder regions indicate a more significant deviation of GQFT from GR (highlighting optimal sky locations for detecting the breathing mode), while bluer regions indicate minimal disparity.}
    \label{skymap}
\end{figure*}

Signals corresponding to both GR and GQFT can be extracted via time--frequency analysis for sources across the sky. Moreover, the detectability of polarization modes depends strongly on the source position, which motivates asking whether certain sky locations are more favorable for discriminating between the two theories. To quantify this distinguishability, we define an amplitude ratio
$\mathrm{Ratio}=\max(\mathrm{Amp})/\min(\mathrm{Amp})$
based on the analysis above. This ratio is chosen to reduce sensitivity to the overall waveform normalization and to characterize a relative intensity. The rate-of-change curves shown in the right panels of Figures~\ref{dif12}--\ref{dif23} help identify extrema and thus facilitate the computation of the amplitude ratio. 
Because the magnitude of this difference varies non-linearly across the sky and spans a large dynamic range, therefore we applied a logarithmic scale to the sky map to visualize these differences more effectively.
Thus we then compute the logarithmic difference between the GR and GQFT ratios $\mathrm{ln} (\mathrm{Ratio}_{\mathrm{GQFT}}-\mathrm{Ratio}_{\mathrm{GR}})$ at each sky position. 
This quantity provides a convenient discriminator and captures the contribution of the $b$ mode. Based on this, we plot an all-sky map as shown in Figure~\ref{skymap}.

Figure~\ref{skymap} shows that only certain sky locations can effectively differentiate between GR and GQFT. The map exhibits several figure-eight structures corresponding to special configurations in which the detector arm is perpendicular to the wavefront and the response vanishes. This occurs because the gravitational waves in both GR and GQFT are transverse. Furthermore, the regions surrounding these structures tend to show larger discrepancies (redder colors), consistent with our earlier finding that the scalar mode is minimally affected by the detector orientation, thereby enhancing the relative contribution of the $b$ mode. In addition, the favorable sky locations are more widely distributed for arm23, suggesting that this arm has higher sensitivity to the $b$ mode and may therefore provide a more promising channel for detecting GQFT signatures. 

Our analysis focuses on distinguishing GR and GQFT through polarization responses, making the details of the source model largely irrelevant. For far-field GWs, the detector response to each polarization is the key ingredient. The GR and GQFT responses appear broadly similar because the $b$, $+$, and $\times$ modes are all transverse, with the $b$ mode exhibiting isotropic transverse breathing. This symmetry makes it nearly impossible to simultaneously null both the $+$ and $\times$ modes, motivating our focus on single-arm variations rather than interferometric combinations. Simple differential combinations of the dominant tensor modes typically leave only a small residual $b$-mode signal, suggesting that single-arm measurements may be more sensitive to this polarization. In future work, we will consider more sophisticated combinations to achieve better separation of the $b$ mode. 

\section{Conclusion}
This work establishes a new methodology for testing quantum gravity through the unique polarization signatures of gravitational waves, moving beyond the constraints of general relativity. By analyzing the gravitational wave polarizations predicted by Gravitational Quantum Field Theory (GQFT), we have pinpointed observable imprints that directly stem from its fundamental departure from the geometric description of gravity.

Our core finding is that GQFT predicts three observable modes---the two standard tensor modes ($+$, $\times$) and a massless breathing scalar mode---while inherently suppressing the vector modes. This distinctive pattern arises not from a mere parametric adjustment, but from the profound theoretical shift in GQFT, where the spin-related gravigauge field, rather than the metric, serves as the fundamental gravitational entity. The physical significance of the vector modes is intrinsically tied to spin-gravity coupling, highlighting the unique role of spin in mediating gravitational interactions within this framework.

To translate this theoretical insight into observable criteria, we developed a practical, model-independent framework. Using first-order orbital dynamics in the Solar System Barycenter frame and the parameterized post-Einsteinian formalism, we derived compact analytical expressions for the detector response in missions like Taiji. This formalism revealed three critical and general observational consequences:
\begin{itemize}
\item \textit{Mode interference}: characteristic interference patterns between the tensor and breathing modes.
\item \textit{Sky-dependent detectability}: the ability to distinguish between GR and GQFT depends strongly on the source sky location, with only certain positions providing optimal discriminatory power.
\item \textit{Arm-specific sensitivity}: different interferometer arms exhibit different sensitivities to the breathing mode, as demonstrated by the distinct response of one arm in our HM Cancri example. This provides a concrete methodology for extracting the scalar polarization signal.
\end{itemize}
Crucially, our polarization mapping technique offers near-complete sky coverage and is fully compatible with existing mission designs, providing a viable and powerful path forward without requiring prohibitively challenging direct measurements.

In summary, this study does more than present a set of calculations; it provides a new paradigm for fundamental physics with future gravitational wave observatories. By focusing on the complete polarization fingerprint, we offer a direct window into the quantum nature of gravity. This approach, generalizable beyond GQFT, equips the next generation of detectors with the theoretical and data-analysis tools needed to probe the ultimate structure of spacetime.

\section*{Acknowledgements}
This work is funded by the National Astronomical Observatories of the Chinese Academy of Sciences (Project No.~E4TG6601). This work has also been supported in part by the National Key Research and Development Program of China (Grant Nos.~2021YFC2203000 and 2020YFC2201501), the National Science Foundation of China (NSFC) (Grant Nos.~12147103, special fund to the center for quanta-to-cosmos theoretical physics, and 11821505), and the Strategic Priority Research Program of the Chinese Academy of Sciences (Grant No.~XDB23030100).
\section*{Conflict of Interest}  The authors declare that they have no conflict of interest.
\end{multicols}
\section*{References}
\bibliographystyle{scichina}
\bibliography{library,references}
\end{document}